\documentclass[preprint,showpacs,preprintnumbers,amsmath,amssymb,aps]{revtex4}
\usepackage{graphicx}
\usepackage{dcolumn}
\usepackage{bm}

\def\lesssim{\ \raise.3ex\hbox{$<$}\kern-0.8em\lower.7ex\hbox{$\sim$}\ }
\def\gesim{\ \raise.3ex\hbox{$>$}\kern-0.8em\lower.7ex\hbox{$\sim$}\ }

\def\rnum#1{\expandafter{\romannumeral #1}} 
\def\Rnum#1{\uppercase\expandafter{\romannumeral #1}} 

\begin{document}
\title{Pairing fluctuations and anisotropic pseudogap phenomenon in an ultracold superfluid Fermi gas with plural $p$-wave superfluid phases}
\author{Daisuke Inotani, and Yoji Ohashi}
\affiliation{Department of Physics, Keio University, 3-14-1 Hiyoshi, Kohoku-ku, Yokohama 223-8522, Japan}
\date{\today}
\begin{abstract}
We investigate superfluid properties of a one-component Fermi gas with a uniaxially anisotropic $p$-wave pairing interaction, $U_x>U_y=U_z$ (where $U_i$ ($i=x,y,z)$ is a $p_i$-wave pairing interaction). This type of interaction is considered to be realized in a $^{40}$K Fermi gas. Including pairing fluctuations within a strong-coupling $T$-matrix theory, we determine the $p_x$-wave superfluid phase transition temperature $T^{p_x}_{\rm c}$, as well as the other phase transition temperature $T_{\rm c}^{p_x+ip_y}$ ($<T_{\rm c}^{p_x}$), below which the superfluid order parameter has the $p_x+ip_y$-wave symmetry. In the normal state near $T^{p_x}_{\rm c}$, $p_x$-wave pairing fluctuations are shown to induce an anisotropic pseudogap phenomenon, where a dip structure in the angle-resolved density of states around $\omega=0$ is the most remarkable in the $p_x$ direction. In the $p_x$-wave superfluid phase ($T_{\rm c}^{p_x+ip_y}<T\le T_{\rm c}^{p_x}$), while the pseudogap in the $p_x$ direction continuously changes to the superfluid gap, the pseudogap in the perpendicular direction to the $p_x$ axis is found to continue developing, because of enhanced $p_y$-wave and $p_z$-wave pairing fluctuations around the node of the $p_x$-wave superfluid order parameter. Since pairing fluctuations are always suppressed in the isotropic $s$-wave superfluid state, this phenomenon is peculiar to an unconventional Fermi superfluid with a nodal superfluid order parameter. Since the $p$-wave Fermi superfluid is the most promising non $s$-wave pairing state in an ultracold Fermi gas, our results would contribute to understanding how the anisotropic pairing fluctuations, as well as the existence of plural superfluid phases, affect many-body properties of this unconventional Fermi superfluid.
\end{abstract}
\pacs{03.75.Ss,05.30.Fk,67.85.-d}
\maketitle
\section{Introduction}
An interesting feature of a non $s$-wave Fermi condensate is that it may have plural superfluid phases, originating from active orbital and/or spin degrees of freedom. Indeed, this possibility has experimentally been confirmed in various Fermi superfluid systems, such as heavy fermion superconductor UPt$_3$ \cite{Sigrist}, as well as superfluid liquid $^3$He \cite{Vollhardt,Mineev}. However, the pairing symmetry that has already been realized is still only the simplest $s$-wave one in cold Fermi gas physics \cite{Jin,Zwierlein,Kinast,Bartenstein}. Thus, going beyond this situation is a crucial challenge in this research field. 
\par
In this regard, a $p$-wave superfluid Fermi gas is a strong candidate, and the possibility of this spin-triplet pairing state has extensively been discussed both experimentally \cite{Regal,Regal2,Ticknor,Zhang,Schunck,Gunter,Gaebler2,Inaba,Fuchs,Mukaiyama,Maier} and theoretically \cite{Gurarie,Ohashi,Ho,Botelho,Iskin,Iskin2,Cheng,Levinsen,Gurarie2,Grosfeld,Iskin3,Mizushima,Han,Mizushima2,Inotani}. Several experimental groups have discovered a $p$-wave Feshbach resonance in $^{40}$K \cite{Regal,Ticknor} and $^6$Li \cite{Zhang,Schunck} Fermi gases, so that we can now tune the strength of a $p$-wave interaction from the weak-coupling regime to the strong-coupling regime, by adjusting an external magnetic field. This experimental development has realized $p$-wave Feshbach molecules \cite{Regal2,Zhang,Gaebler2,Inaba,Fuchs,Mukaiyama}. Thus, although one still needs to overcome some difficulties, such as the three-body loss \cite{Levinsen,Castin,Gurarie3}, as well as the dipolar relaxation \cite{Zhang,Gaebler2} (that destroy $p$-wave molecules \cite{Chevy}), the $p$-wave superfluid state seems a very promising non $s$-wave pairing state in an ultracold Fermi gas.
\par
It has been predicted \cite{Gurarie,Gurarie2} that, when a $p$-wave interaction has a uniaxial anisotropy, a $p$-wave superfluid Fermi gas may have two superfluid phases with different $p$-wave pairing symmetries. Such an anisotropic $p$-wave pairing interaction is considered to be realized in a $^{40}$K Fermi gas \cite{Ticknor}, because the split of a $p$-wave Feshbach resonance into a $p_x$-wave channel and degenerate $p_y$-wave and $p_z$-wave channels by a magnetic dipole-dipole interaction has been observed, when an external magnetic field is applied in the $x$ direction. Since the observed resonance field of a $p_x$-wave Feshbach resonance is higher than that of the other degenerate channels \cite{Ticknor}, a $p$-wave pairing interaction associated with this $p$-wave Feshbach resonance has the uniaxial anisotropy, that is, a $p_x$-wave pairing interaction $U_x$ is stronger than $p_y$-wave ($U_y$) and $p_z$-wave ($U_z$) interactions. As a result, the $p_x$-wave superfluid state has the highest superfluid phase transition temperature $T_{\rm c}^{p_x}$. In this case, Refs. \cite{Gurarie,Gurarie2} have pointed out that the system experiences the other phase transition from the $p_x$-wave state to the $p_x+ip_y$-wave state at $T_{\rm c}^{p_x+ip_y}$ ($<T_{\rm c}^{p_x}$), when the uniaxial anisotropy satisfies a certain condition. Thus, the realization of a $p$-wave superfluid Fermi gas would enable us to study physics of plural superfluid phases, from the weak-coupling regime to the strong-coupling limit in a systematic manner.
\par
When the above-mentioned $p$-wave superfluid Fermi gas is realized, the existence of strong pairing fluctuations near $T_{\rm c}^{p_x+ip_y}$ is an interesting research topic. In the $p_x$-wave superfluid phase, since single-particle excitations are gapless in the nodal direction ($\perp p_x$) of the $p_x$-wave superfluid order parameter $\Delta_{p_x}({\bm p})\propto p_x$, $p_y$-wave and $p_z$-wave pairing fluctuations can continue developing, to be the strongest at $T_{\rm c}^{p_x+ip_y}$. Thus, even far below $T_{\rm c}^{p_x}$, an anisotropic pseudogap phenomenon is expected in the nodal direction of the $p_x$-wave superfluid order parameter near $T_{\rm c}^{p_x+ip_y}$. 
\par
In an isotropic $s$-wave superfluid Fermi gas, since the BCS gap opens in all the momentum direction of single-particle excitations, pairing fluctuations are soon suppressed below the superfluid phase transition temperature $T_{\rm c}$. Indeed, it has been shown that a pseudogap in the density of states above $T_{\rm c}$ \cite{Stewart,Gaebler,Perali2,Perali,Tsuchiya1,Levin,Hui,Bulgac} soon changes to the $s$-wave BCS superfluid gap below $T_{\rm c}$ \cite{Watanabe}. Even in the $p_x$-wave case, the enhancement of pairing fluctuations around the node would not occur in the absence of a $p_y$-wave and a $p_z$-wave interactions, because $p_x$-wave pairing fluctuations are soon suppressed by the $p_x$-wave superfluid order below $T_{\rm c}^{p_x}$. Thus, the above-mentioned pseudogap phenomenon in the $p_x$-wave state near $T_{\rm c}^{p_x+ip_y}$ is peculiar to an unconventional Fermi superfluid with a nodal superfluid order parameter, as well as with plural superfluid phases. 
\par
In this paper, we theoretically investigate strong-coupling properties of a one component $p$-wave superfluid Fermi gas with a uniaxially anisotropic $p$-wave pairing interaction ($U_x>U_y=U_z$). Including $p$-wave pairing fluctuations within the framework of a strong coupling $T$-matrix approximation, we determine $T_{\rm c}^{p_x}$ and $T_{\rm c}^{p_x+ip_y}$. In the $p_x$-wave superfluid phase, we calculate the angle-resolved single-particle density of states, to clarify that pairing fluctuations in the nodal direction ($\perp p_x$) of the $p_x$-wave superfluid order parameter continue to develop, leading to a pseudogapped single-particle excitation spectrum in the nodal direction. 
\par
The paper is organized as follows. In Sec.II, we explain our strong-coupling formalism for a $p$-wave superfluid Fermi gas. In Sec.III, we show our numerical results on $T_{\rm c}^{p_x}$ and $T_{\rm c}^{p_x+ip_y}$. Here, we clarify the condition for the appearance of $p_x$-wave and $p_x+ip_y$-wave superfluid phases. In Sec.IV, we examine the angle-resolved single-particle density of states (ARDOS). We show that a pseudogap anisotropically appears in ARDOS, not only in the normal state near $T_{\rm c}^{p_x}$, but also in the $p_x$-wave superfluid phase near $T_{\rm c}^{p_x+ip_y}$. To characterize this anisotropic many-body phenomenon, we introduce the characteristic temperature $T^*$ as the temperature below which a dip structure appears in ARDOS. Throughout this paper, we take $\hbar=k_{B}=1$, and the volume of the system $V$ is taken to be unity, for simplicity.
\par
\section{Formulation}
\par
We consider a one-component Fermi gas with a uniaxially anisotropic $p$-wave interaction, described by the Hamiltonian,
\begin{equation}
H=\sum_{\bm p} \xi_{\bm p}c_{\bm p}^{\dagger}c_{\bm p}
-\frac{1}{2}\sum_{{\bm p},{\bm p}',{\bm q}} 
\sum_{i=x,y,z} F_n({\bm p}) p_i U_i p'_i F_n({\bm p}')
c_{{\bm p}+{\bm q}/2}^\dagger c_{-{\bm p}+{\bm q}/2}^\dagger
c_{-{\bm p}'+{\bm q}/2}c_{{\bm p}'+{\bm q}/2}.
\label{eq.1}
\end{equation}
Here, $c^{\dagger}_{\bm p}$ is the creation operator of a Fermi atom with the kinetic energy $\xi_{\bm p}=\varepsilon_{\bm p}-\mu=p^2/2m-\mu$, measured from the Fermi chemical potential $\mu$ (where $m$ is the atomic mass). $-U_i$ ($<0$) is a pairing interaction in the $p_i$-wave Cooper channel ($i=x,y,z$), having the uniaxial anisotropy $U_x > U_y= U_z$ \cite{Ticknor}. In this paper, we do not deal with details of a $p$-wave Feshbach resonance, but simply treat $(U_x,U_y,U_z)$ as a tunable parameter set. To eliminate the ultraviolet divergence, the last term in Eq. (\ref{eq.1}) involves the cutoff function \cite{Ohashi,Ho,Botelho,Iskin2,Inotani}, 
\begin{equation}
F_n({\bm p})={1 \over 1+(p/p_{\rm c})^{2n}},  
\label{eq.2}
\end{equation}
where $p_{\rm c}$ is a cutoff momentum. For simplicity, we use the same cutoff function $F_n({\bm p})$ for all the $p$-wave Cooper channels. Equation (\ref{eq.2}) gives a Lorentzian cutoff when $n=1$ \cite{Iskin2}, and gives a sharp cutoff when $n=\infty$ \cite{Inotani}.  We will discuss the cutoff dependence of our results in Sec. III.
\par
As usual \cite{Ohashi,Ho,Botelho,Iskin2,Inotani}, we conveniently measure the strength of a $p$-wave interaction in terms of the $p_i$-wave scattering volume $v_i$ ($i=x,y,z$), as well as the effective range $k_0$, that are given by, respectively,
\begin{equation}
{4\pi v_i \over m}=-
{U_i \over 3}
{1 \over \displaystyle 1-{U_i \over 3}\sum_{\bm p}{p^2 \over 2\varepsilon_{\bm p}}F_n^2({\bm p})},
\label{eq.3}
\end{equation}
\begin{equation}
k_0=-{4\pi \over m^2}
\sum_{\bm p}{p^2 \over 2\varepsilon_{\bm p}^2}F_n^2({\bm p}).
\label{eq.4}
\end{equation}
Since we take the same cutoff function $F_n({\bm p})$ for all the $p_i$-wave interaction channels, the effective range $k_0$ are channel-independent. We also introduce the anisotropy parameter, 
\begin{equation}
\delta v^{-1}\equiv v^{-1}_x-v^{-1}_y=v^{-1}_x-v^{-1}_z~(>0).
\label{eq.5b}
\end{equation}
Then, the $p$-wave interaction can be specified by the parameter set $(v_x^{-1}, \delta v^{-1}, k_0, n)$. The weak-coupling side and the strong-coupling side are characterized as $(p_{\rm F}^3v_x)^{-1}\lesssim 0$ and $(p_{\rm F}^3v_x)^{-1}\gesim 0$, respectively, where $p_{\rm F}$ is the Fermi momentum. 
\par
To deal with strong-coupling phenomena in the superfluid phase, it is convenient to rewrite the model Hamiltonian in Eq. (\ref{eq.1}) into the sum of the mean-field BCS part and the term describing fluctuation corrections \cite{Watanabe,OhashiTakada}. Under the Nambu representation \cite{Schrieffer}, we have,
\begin{eqnarray}
H=\frac{1}{2}\sum_{\bm p} \Psi_{\bm p}^{\dagger}
\left[ \xi_{\bm p} \tau_3 - \hat{\Delta}({\bm p}) \right] \Psi_{\bm p}
-\frac{1}{2}\sum_{{\bm q},i=x,y,z} U_i\rho_i^+({\bm q})\rho_i^-(-{\bm q}),
\label{eq.7}
\end{eqnarray}
where 
\begin{eqnarray}
\Psi_{\bm p}=\left(
\begin{array}{c}
c_{\bm p}   \\
c_{-{\bm p}} ^{\dagger}   
\end{array}
\right)
\label{eq.6}
\end{eqnarray}
is the two-component Nambu field, and $\tau_i$ ($i=x,y,z$) are Pauli matrices acting on particle-hole space \cite{Schrieffer}. The first term in Eq. (\ref{eq.7}) is just the mean-field BCS Hamiltonian, where
\begin{eqnarray}
\hat{\Delta}({\bm p})=
\left(
\begin{array}{cc}
0 & \Delta({\bm p}) \\
\Delta^*({\bm p}) & 0
\end{array}
\right)
\label{eq.6b}
\end{eqnarray}
is a $2\times 2$ matrix $p$-wave superfluid order parameter. Here,
\begin{eqnarray}
\Delta({\bm p})={\bm b} \cdot {\bm p}F_n({\bm p}),
\label{eq.8}
\end{eqnarray}
where ${\bm b}=(b_x,b_y,b_z)$ has the form,
\begin{equation}
b_i=U_i\sum_{\bm p}p_iF_n({\bm p})
\langle c_{-{\bm p}}c_{{\bm p}} \rangle. 
\label{eq.8b}
\end{equation}
The second term in Eq. (\ref{eq.7}) gives fluctuation corrections to the mean-field Hamiltonian, where the so-called generalized density operator \cite{Watanabe,OhashiTakada},
\begin{eqnarray}
\rho_i^\pm ({\bm q})= \sum_{\bm p} p_i F_n({\bm p}) \Psi_{{\bm p} +\frac{\bm q}{2}}^{\dagger} \tau_\pm \Psi_{{\bm p} -\frac{\bm q}{2}},
\label{eq.9}
\end{eqnarray}
physically describes $p_i$-wave superfluid fluctuations (where $\tau_\pm=\left( \tau_1\pm i\tau_2 \right)/2$).
\par
\begin{figure}
\centerline{\includegraphics[width=10cm]{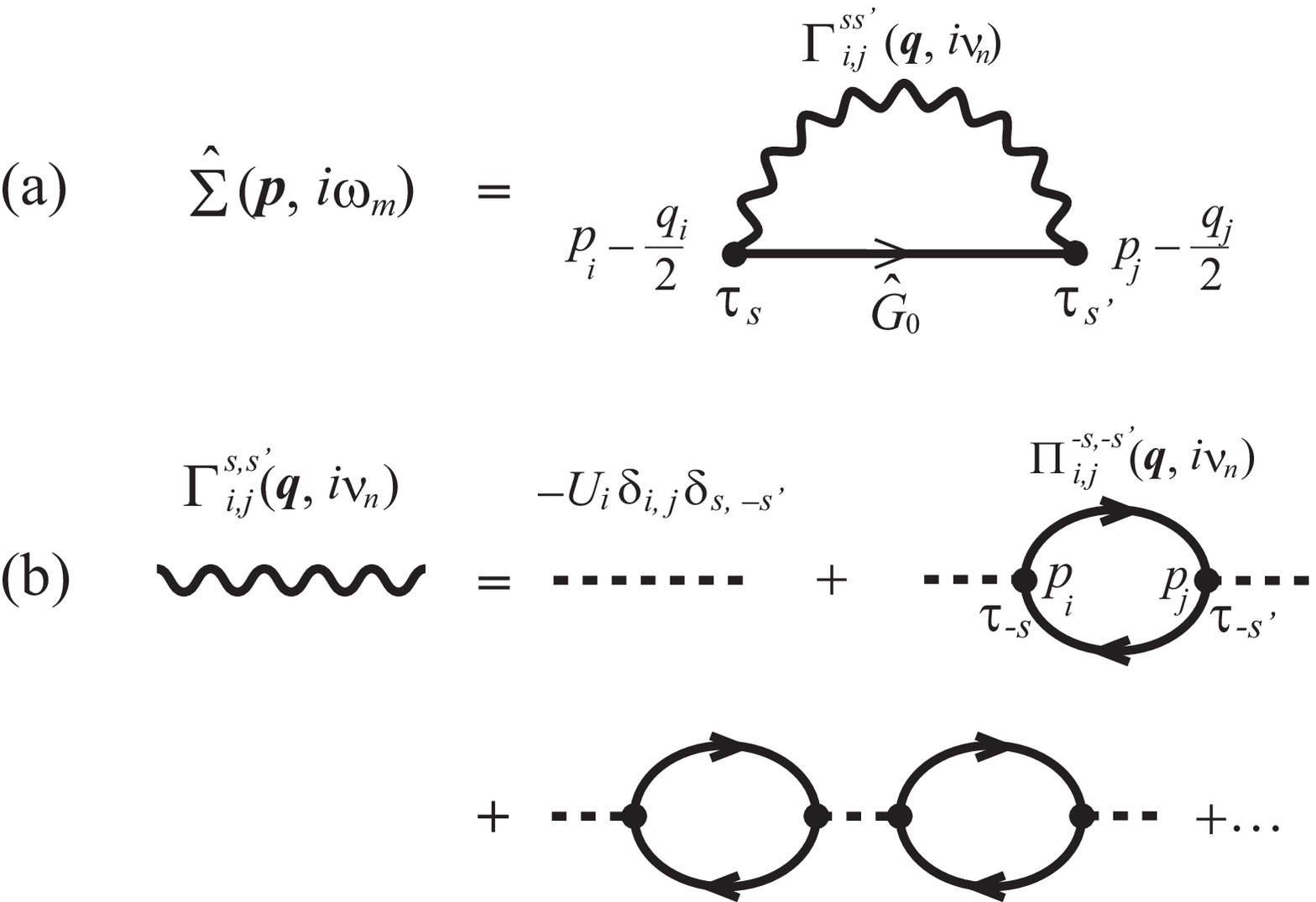}}
\caption{(a) Self-energy correction $\hat{\Sigma}({\bm p}, i\omega_n)$ in the $T$-matrix approximation (TMA). The wavy line is the particle-particle scattering matrix $\Gamma_{i,j}^{s,s'}({\bm q}, i\nu_n)$ given in (b). The solid line and the dashed line describe the mean field single-particle Green's function $\hat{G}_0({\bm p},i \omega_n)$, and the $p$-wave interaction, respectively. In (a), the factor $(p_i-q_i/2)F_n({\bm p}-{\bm q}/2)\tau_s$ is assigned to each vertex (solid circle). In (b), we assign $p_iF_n({\bm p})\tau_s$ to each vertex. In this figure, $-s$ means the opposite sign to $s=\pm$.}
\label{fig1}
\end{figure}
\par
Strong-coupling corrections to single-particle excitations can be conveniently described by the self-energy $\hat{\Sigma}({\bm p},i\omega_n)$ in the single-particle thermal Green's function,
\begin{eqnarray}
\hat{G}({\bm p}, i\omega_m)=\frac{1}{\hat{G}_0^{-1}\left({\bm p}, i\omega_m \right)-\hat{\Sigma}\left({\bm p}, i\omega_m \right)},
\label{eq.10}
\end{eqnarray}
where $\omega_m$ is the fermion Matsubara frequency. In Eq. (\ref{eq.10}), 
\begin{eqnarray}
\hat{G}_0({\bm p},i\omega_m)=\frac{1}{i\omega_m-\xi_{\bm p} \tau_3
+\hat{\Delta}({\bm p})}
\label{eq.11}
\end{eqnarray}
is the single-particle Green's function in the mean-field level. Treating the last term in Eq. (\ref{eq.7}) within the $T$-matrix approximation (TMA) \cite{Perali,Tsuchiya1,Watanabe}, we have the diagrammatic expression for the self-energy $\hat{\Sigma}({\bm p},i\omega_n)$ shown in Fig. \ref{fig1}(a) \cite{note1}, which gives
\begin{eqnarray}
{\hat \Sigma}({\bm p},i\omega_m)
&=&
{2 \over \beta}
\sum_{i,j=x,y,z}
\sum_{{\bm q},i\nu_n}
F_n^2\left({\bm p}-\frac{{\bm q}}{2}\right)
\left[p_i-{q_i \over 2}\right]
\left[p_j-{q_j \over 2}\right]
\nonumber
\\
&\times&
\left(
\begin{array}{cc}
G_0^{22}({\bm p}-{\bm q},i\omega_n-i\nu_n) 
\Gamma^{-+}_{i,j}({\bm q},i\nu_n) &
G_0^{21}({\bm p}-{\bm q},i\omega_n-i\nu_n)
\Gamma^{++}_{i,j}({\bm q},i\nu_n) \\
G_0^{12}({\bm p}-{\bm q},i\omega_n-i\nu_n)
\Gamma^{--}_{i,j}({\bm q},i\nu_n) &
G_0^{11}({\bm p}-{\bm q},i\omega_n-i\nu_n)
\Gamma^{+-}_{i,j}({\bm q},i\nu_n)
\end{array}
\right).
\nonumber
\\
\label{eq.12}
\end{eqnarray}
Here, $\nu_n$ is the boson Matsubara frequency. The TMA particle-particle scattering matrix $\Gamma_{i,j}^{s,s'}({\bm q},i\nu_n)$ in Eq. (\ref{eq.12}) obeys the equation,
\begin{eqnarray}
\Gamma_{i,j}^{s,s'}({\bm q},i\nu_n)=-U_i \delta_{i,j}\delta_{s,-s'}
-U_i \sum_{s''=\pm}\sum_{k=x,y,z} 
\Pi_{i,k}^{s, s''}({\bm q},i\nu_n)
\Gamma_{k,j}^{-s'', s'}({\bm q},i\nu_n),
\label{eq.14}
\end{eqnarray}
where $-s$ means the opposite sign to $s=\pm$, and 
\begin{eqnarray}
\Pi_{i,j}^{s,s'}({\bm q},i\nu_n)=
\frac{1}{\beta}\sum_{\bm p} p_ip_j F_n^2(p){\rm Tr} \left[
\tau_s \hat{G}_0\left({\bm p}+\frac{\bm q}{2},i\omega_n \right) 
\tau_{s'} \hat{G}_0\left( {\bm p}-\frac{\bm q}{2},i\omega_n-i\nu_n \right)
\right]
\label{eq.15}
\end{eqnarray}
is the pair correlation function. In particular, $\Pi_{i,i}^{s,s'}({\bm q},i\nu_n)$ describes fluctuations in the $p_i$-wave Cooper channel, and $\Pi_{i,j}^{s,s'}({\bm q},i\nu_n)$ ($i\ne j$) describes coupling between $p_i$-wave and $p_j$-wave pairing fluctuations. 
\par
In the present anisotropic case ($U_x>U_y=U_z$), the highest superfluid phase transition temperature is obtained in the $p_x$-wave Cooper channel. As in the $s$-wave case \cite{Watanabe,Thouless}, the TMA gap equation for the $p_x$-wave superfluid order parameter $\Delta_{p_x}({\bm p})=b_xp_xF_n({\bm p})$ is obtained from the Thouless criterion $\left[ \Gamma_{x,x}^{{\rm ph}}({\bm q}=0,i\nu_n=0)\right]^{-1}=0$ in the $p_x$-wave Cooper channel, where
\begin{eqnarray}
\Gamma_{i,j}^{{\rm ph}}({\bm q},i\nu_n)=\frac{1}{4}
\left[
\Gamma_{i,j}^{-+}({\bm q},i\nu_n)
+\Gamma_{i,j}^{+-}({\bm q},i\nu_n)
-\Gamma_{i,j}^{--}({\bm q},i\nu_n)
-\Gamma_{i,j}^{++}({\bm q},i\nu_n)
\right],
\label{eq.14_2}
\end{eqnarray}
describes the phase fluctuations of the $p_x$-wave superfluid order parameter  $\Delta_{p_x}({\bm p})$. Physically, this guarantees the existence of a gapless Goldstone mode associated with the broken $U(1)$ gauge symmetry. Noting that $b_y=b_z=0$ in the $p_x$-wave superfluid phase, we obtain the gap equation in the $p_x$-wave superfluid state as
\begin{eqnarray}
1={12\pi v_x \over m}\sum_{\bm p}p_x^2F_n^2({\bm p})
\left[
{1 \over 2\sqrt{\xi_{\bm p}^2+|\Delta_{p_x}({\bm p})|^2}}
\tanh{\sqrt{\xi_{\bm p}^2+|\Delta_{p_x}({\bm p})|^2} \over 2T}
-
{1 \over 2\varepsilon_{\bm p}}
\right].
\label{eq.18q}
\end{eqnarray}
The equation for $T_{\rm c}^{p_x}$ is obtained from Eq. (\ref{eq.18q}) by setting $\Delta_{p_x}({\bm p})=0$ as
\begin{eqnarray}
1={12\pi v_x \over m}\sum_{\bm p}p_x^2F_n^2({\bm p})
\left[
{1 \over 2\xi_{\bm p}}\tanh{\xi_{\bm p} \over 2T_{\rm c}^{p_x}}
-
{1 \over 2\varepsilon_{\bm p}}
\right].
\label{eq.18}
\end{eqnarray}
\par
In the case of uniaxially anisotropic $p$-wave interaction, Ref. \cite{Gurarie2} pointed out that, without loss of generality, one may restrict the structure of the $p$-wave superfluid order parameter to the form,
\begin{eqnarray}
{\bm b}=
\left(
\begin{array}{c}
b_x\\
b_y\\
b_z\\
\end{array}
\right)
=
\left(
\begin{array}{c}
B_x\\
iB_y\\
0\\
\end{array}
\right),
\label{eq.18b1}
\end{eqnarray}
where $B_x$ and $B_y$ are real quantities. Thus, in the $p_x$-wave superfluid phase below $T_{\rm c}^{p_x}$ (where $B_x=b_x\ne0$ and $B_y=0$), the other possible superfluid instability is only associated with the $p_x+ip_y$-wave one, having the superfluid order parameter, 
\begin{eqnarray}
\Delta_{p_x+ip_y}({\bm p})=[B_xp_x+iB_yp_y]F_n({\bm p}).
\label{eq.18b}
\end{eqnarray}
Since $B_x$ is already present below $T_{\rm c}^{p_x}$, the superfluid phase transition temperature $T_{\rm c}^{p_x+ip_y}$ is determined from the Thouless criterion \cite{Thouless} in the $p_y$-wave Cooper channel $\left[ \Gamma_{y,y}^{{\rm ph}}({\bm q}=0,i\nu_n=0)\right]^{-1}=0$, 
\begin{eqnarray}
1={12\pi v_y \over m}\sum_{\bm p}p_y^2F_n^2({\bm p})
\left[
{1 \over 2\sqrt{\xi_{\bm p}^2+|\Delta_{p_x}({\bm p})|^2}}
\tanh{\sqrt{\xi_{\bm p}^2+|\Delta_{p_x}({\bm p})|^2} \over 2T_{\rm c}^{p_x+ip_y}}
-
{1 \over 2\varepsilon_{\bm p}}
\right],
\label{eq.18c}
\end{eqnarray}
where the $p_x$-wave superfluid order parameter $\Delta_{p_x}({\bm p})=b_xp_xF_n({\bm p})$ obeys the gap equation (\ref{eq.18q}). 
\par
We numerically solve Eqs. (\ref{eq.18q}), (\ref{eq.18}), and (\ref{eq.18c}), to self-consistently determine $T_{\rm c}^{p_x}$, $T_{\rm c}^{p_x+ip_y}$, and $\Delta_{p_x}({\bm p})$. In this procedure, we also solve the equation for the total number $N_{\rm F}$ of Fermi atoms,
\begin{equation}
N_{\rm F}={T \over 2}\sum_{{\bm p},i\omega_n}
{\rm Tr}
[\tau_3{\hat G}({\bm p},i\omega_n)],
\label{eq.18d}
\end{equation}
to include strong-coupling corrections to the Fermi chemical potential $\mu$. 
\par
We examine the anisotropic pseudogap phenomenon by calculating the angle-resolved single-particle density of states (ARDOS), 
\begin{equation}
\rho(\omega,\hat{\bm p})=
-{1 \over \pi}
\int_0^\infty {p^2 dp \over (2\pi)^3}
{\rm Im } \left[ 
G_{11}({\bm p},i\omega_n \to \omega + i\delta)
\right],
\label{eq.27}
\end{equation}
where $\hat{\bm p}={\bm p}/|{\bm p}|$, and $G_{11}({\bm p},i\omega_n \to \omega + i\delta)$ is the (1,1) component of the analytic continued TMA Green's function in Eq. (\ref{eq.10}). ARDOS in Eq. (\ref{eq.27}) is related to the ordinary density of states $\rho(\omega)$ as
\begin{equation}
\rho(\omega)=\int \sin\theta_{\bm p}d\theta_{\bm p}d\phi_{\bm p}
\rho(\omega,\hat{\bm p}).
\label{eq.27b}
\end{equation}
Here, we choose the $p_x$ axis as the polar axis ($p_x=p \cos \theta_{\bm p}$).
\par
We note that, since the anisotropy of the $p_x$-wave superfluid order parameter $\Delta_{p_x}({\bm p})\propto p_x$ lowers the symmetry of the system, we need much time to compute the number equation (\ref{eq.18d}) below $T_{\rm c}^{p_x}$, compared to the symmetric $s$-wave case. To avoid this difficulty, in this paper, we approximate the TMA Green's function ${\hat G}({\bm p},i\omega_n)$ in the number equation (\ref{eq.18d}) to
\begin{equation}
{\hat G}({\bm p},i\omega_n)\simeq{\hat G}_0({\bm p},i\omega_n)+
{\hat G}_0({\bm p},i\omega_n){\hat \Sigma}({\bm p},i\omega_n){\hat G}_0
({\bm p},i\omega_n).
\label{eq.18e}
\end{equation}
Equation (\ref{eq.18e}) is just the same form as the Green's function in the strong-coupling theory developed by Nozi\`eres and Schmitt-Rink (NSR) \cite{NSR}. The NSR theory has extensively been used in the $s$-wave case, to successfully explain the BCS-BEC crossover behavior of the superfluid phase transition temperature \cite{NSR,Melo}, as well as the superfluid order parameter in the crossover region \cite{Randeria,OhashiGriffin}. The NSR theory has also been extended to the $p$-wave case with $U_x=U_y=U_z$ \cite{Ohashi}. Thus, we expect that the NSR Green's function in Eq. (\ref{eq.18e}) also works in determining $T_{\rm c}^{p_x}$, $T_{\rm c}^{p_x+ip_y}$, $\mu$ and $\Delta_{p_x}({\bm p})$.
\par
On the other hand, it is also known that the NSR theory unphysically gives negative density of states in the BCS-BEC crossover region \cite{Tsuchiya1,Kashimura2012}. Since this serious problem is absent in TMA, we use the TMA Green's function in Eq. (\ref{eq.10}), in considering single-particle properties of a $p_x$-wave Fermi superfluid. 
\par
Here, we summarize our detailed parameter settings. For the effective range, we take $k_0=-30p_{\rm F}$, following the experimental result on a $^{40}$K Fermi gas \cite{Ticknor}. For the cutoff function $F_n({\bm p})$ in Eq. (\ref{eq.2}), we set $n=3$. The cutoff momentum $p_{\rm c}$ in $F_n({\bm p})$ is determined so as to reproduce $k_0=-30p_{\rm F}$, which gives $p_{\rm c}=27p_{\rm F}$. Since we only deal with the normal state, as well as the $p_x$-wave superfluid state, ARDOS in Eq. (\ref{eq.27}) is actually independent of the angle $\phi_{\bm p}$ around the $p_x$ axis. Thus, the anisotropy can be simply specified by the polar angle $\theta_{\bm p}$ measured from the $p_x$ axis. Noting this, we write Eq. (\ref{eq.27}) as $\rho(\omega,\theta_{\bm p})$ in what follows.
\par
\section{Phase diagram of an ultracold Fermi gas with $p$-wave interaction}
\par
\begin{figure}
\centerline{\includegraphics[width=10cm]{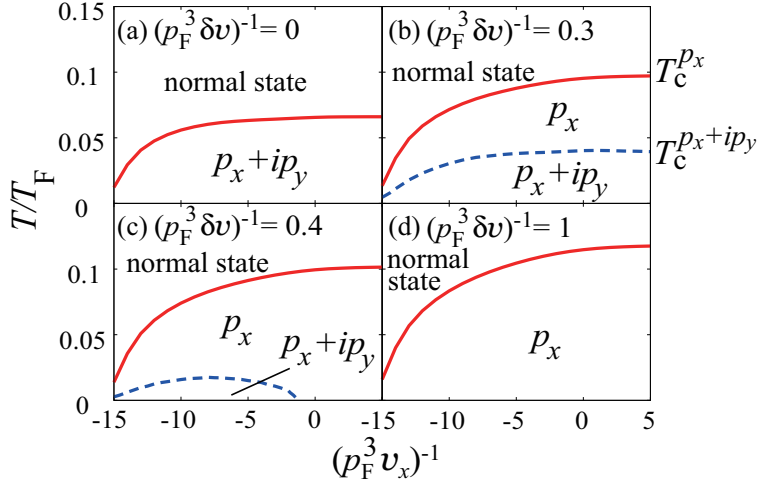}}
\caption{(Color online) Phase diagram of a one-component Fermi gas with a uniaxially anisotropic $p$-wave pairing interaction. The solid line and the dashed line are $T_{\rm c}^{p_x}$ and $T_{\rm c}^{p_x+ip_y}$, respectively. $T_{\rm F}$ is the Fermi temperature.}
\label{fig2}
\end{figure}

Figure \ref{fig2} shows the phase diagram of a one component ultracold Fermi gas in terms of the $p$-wave interaction strength, $(p_{\rm F}^3v_x)^{-1}$, and the temperature. When $\delta v^{-1}=0$ ($U_x=U_y=U_z)$, Fig. \ref{fig2}(a) shows that the superfluid phase is dominated by the $p_x+ip_y$-wave pairing state. This superfluid region gradually shrinks with increasing the uniaxial anisotropy $\delta v^{-1}$, as shown in Figs. \ref{fig2}(b) and (c). Since the present anisotropy ($U_x>U_y=U_z$) favors the $p_x$-wave symmetry, the region of the $p_x+ip_y$-wave state eventually vanishes, as shown in Fig. \ref{fig2}(d). We briefly note that the overall structure of this phase diagram is consistent with the previous work based on mean-field analyses \cite{Gurarie,Gurarie2}.
\par
\par
\begin{figure}
\centerline{\includegraphics[width=10cm]{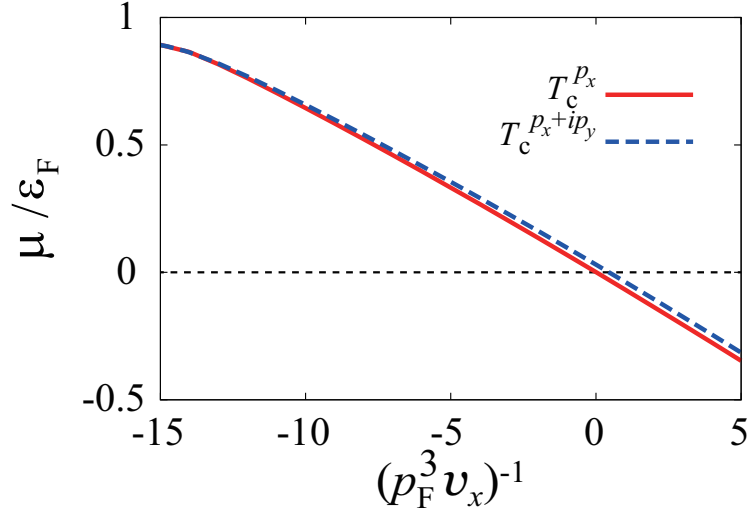}}
\caption{(Color online) Calculated Fermi chemical potential $\mu$ at the two superfluid phase transition temperatures, $T_{\rm c}^{p_x}$ and $T_{\rm c}^{p_x+ip_y}$. We set $(p_{\rm F}^3\delta v)^{-1}=0.3$.}
\label{fig3}
\end{figure}

In addition to $T_{\rm c}^{p_x}$, the coupled equations (\ref{eq.18}) and (\ref{eq.18d}) also give the Fermi chemical potential $\mu(T=T_{\rm c}^{p_x})$ shown in Fig. \ref{fig3}, which exhibits the typical BCS-BEC crossover behavior \cite{Ohashi,Ho,Inotani,NSR,Tsuchiya1,Watanabe,Perali,Melo,Randeria}. That is, with increasing the interaction strength, $\mu(T=T_{\rm c}^{p_x})$ gradually deviates from the Fermi energy $\varepsilon_{\rm F}$, to be negative in the strong-coupling regime, when $(p_{\rm F}^3v_x)^{-1}\gesim 0$. 
\par
\begin{figure}
\centerline{\includegraphics[width=10cm]{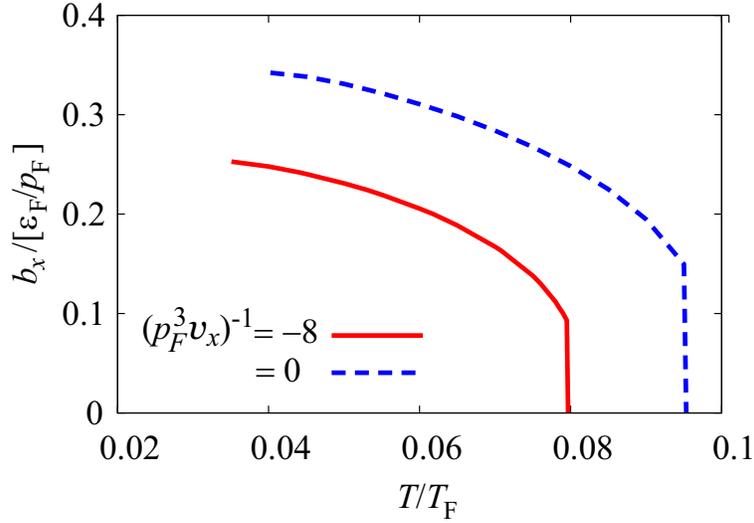}}
\caption{(Color online) Calculated factor $b_x$ in the $p_x$-wave superfluid order parameter $\Delta_{p_x}=b_xp_xF_{n=3}({\bm p})$, when $(p_{\rm F}^3\delta v)^{-1}=0.3$. Each result ends at $T_{\rm c}^{p_x+ip_y}$.}
\label{fig4}
\end{figure}

At $T_{\rm c}^{p_x+ip_y}$, the coupled equations (\ref{eq.18q}), (\ref{eq.18c}) and (\ref{eq.18d}) also give $\mu(T=T_{\rm c}^{p_x+ip_y})$, as well as the $p_x$-wave superfluid order parameter $\Delta_{p_x}({\bm p},T=T_{\rm c}^{p_x+ip_y})$. Figure \ref{fig3} shows that $\mu(T=T_{\rm c}^{p_x+ip_y})\simeq \mu(T=T_{\rm c}^{p_x})$ in the whole interaction regime, indicating that the chemical potential is almost $T$-independent in the $p_x$-wave superfluid phase. For the $p_x$-wave superfluid order parameter $\Delta_{p_x}({\bm p},T=T_{\rm c}^{p_x+ip_y})$, of course, this quantity has already existed above $T_{\rm c}^{p_x+ip_y}$, as shown in Fig. \ref{fig4}. Although the pairing symmetry changes from the $p_x$-wave one to the $p_x+ip_y$-wave one at $T_{\rm c}^{p_x+ip_y}$, it is known \cite{Gurarie2} that this symmetry change occurs smoothly, in the sense that $B_y$ in Eq. (\ref{eq.18b}) continuously grows from zero below $T_{\rm c}^{p_x+ip_y}$. Thus, the second order phase transition is expected at $T_{\rm c}^{p_x+ip_y}$ (unless the superfluid order parameter exhibits an unexpected discontinuity at $T_{\rm c}^{p_x+ip_y}$).
\par
\par
\par
\par
In Fig. \ref{fig4}, we see that $\Delta_{p_x}({\bm p})$ has a discontinuity at $T_{\rm c}^{p_x}$, which is, however, an artifact of TMA we are using in this paper. The same problem has already been known in the $s$-wave case \cite{Watanabe,OhashiGriffin,OhashiJPSJ}. In the latter case, it has been pointed out \cite{OhashiJPSJ} that one needs to correctly include an effective repulsive interaction between Cooper pairs beyond TMA, in order to recover the expected second order phase transition. Although this improvement is also crucial in the $p$-wave case, we leave this problem as a future problem, and examine strong-coupling effects in the $p_x$-wave superfluid phase within TMA. 
\par
\begin{figure}
\centerline{\includegraphics[width=15cm]{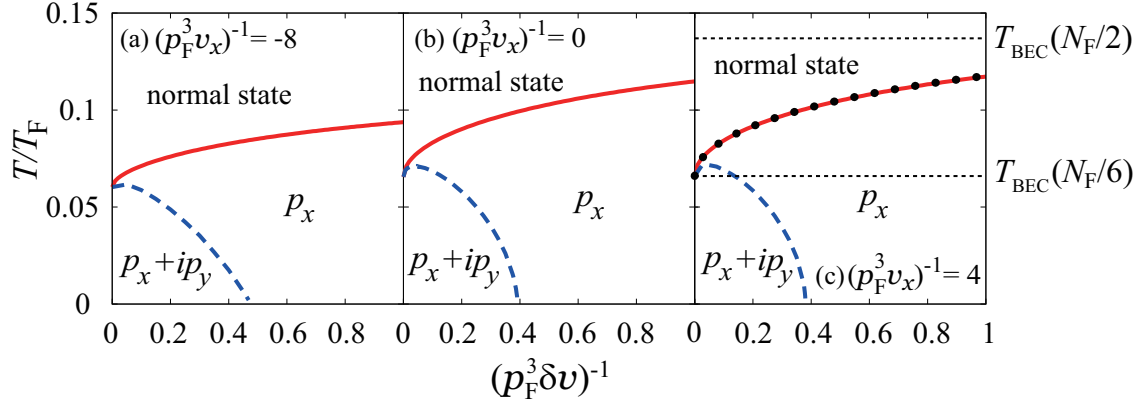}}
\caption{(Color online) Calculated two superfluid phase transition temperatures, $T_{\rm c}^{p_x}$ and $T_{p_x+ip_y}$, as functions of the anisotropy parameter $(p_{\rm F}^3\delta v)^{-1}$. In panel (c), the solid circles shows the BEC phase transition temperature $T_{\rm BEC}$ obtained from Eqs. (\ref{eq.24a})-(\ref{eq.25}). The upper dashed line in this panel shows $T_{\rm BEC}(N_{\rm F}/2)$. The lower dashed line shows $T_{\rm BEC}(N_{\rm F}/6)$.}
\label{fig5}
\end{figure}
\par
In the weak-coupling BCS limit, the number equation (\ref{eq.18d}) simply gives $\mu=\varepsilon_{\rm F}$, so that the superfluid phase transition temperature $T_{\rm c}^{p_x}$ is determined from Eq. (\ref{eq.18}) with $\mu=\varepsilon_{\rm F}$. The resulting $T_{\rm c}^{p_x}$ does not depend on the anisotropy parameter $\delta v^{-1}$, for a fixed $p_x$-wave interaction strength $v_x$. On the other hand, $T_{\rm c}^{p_x}$ gradually comes to depend on $\delta v^{-1}$, as one approaches the strong-coupling regime, as shown in Fig. \ref{fig5}. 
\par
To explain the anisotropy dependence of $T_{\rm c}^{p_x}$ in the strong-coupling regime shown in Fig. \ref{fig5}(c), we first note that the system in the strong-coupling limit \cite{note2} may be viewed as an ideal gas mixture of three kinds of tightly-bound molecules (with the molecular mass $M=2m$) that are formed by the three $p_i$-wave interactions ($i=x,y,z$). Indeed, in the strong-coupling limit, the number equation (\ref{eq.18d}) at $T_{\rm c}^{p_x}$ is reduced to
\begin{equation}
{N_{\rm F} \over 2}=\sum_{i=x,y,z}N_{\rm B}^i,
\label{eq.24a}
\end{equation}
where 
\begin{equation}
N_{\rm B}^i=\sum_{\bm q}n_{\rm B}
\left( \frac{q^2}{2M} -\mu_{\rm B}^i \right)
\label{eq.24b}
\end{equation}
is the number of molecules in the $p_i$-wave Cooper channel, with $n_{\rm B}(\omega)$ being the Bose distribution function. The $T_{\rm c}^{p_x}$-equation (\ref{eq.18}) gives the Bose chemical potential $\mu_{\rm B}^i$ in Eq. (\ref{eq.24b}) as
\begin{eqnarray}
\left\{
\begin{array}{l}
\displaystyle
\mu_{\rm B}^x=0,\\
\displaystyle
\mu_{\rm B}^y=\mu_B^z=-\frac{2\delta v^{-1}}{m \left(|k_0|-3\sqrt{2m|\mu|} \right)}~~~(\le 0).
\end{array}
\right.
\label{eq.25}
\end{eqnarray}
In the absence of uniaxial anisotropy ($\delta v^{-1}$=0), all the three components simultaneously satisfy the BEC condition, $\mu_{\rm B}^x=\mu_{\rm B}^y=\mu_{\rm B}^z=0$. Thus, the phase transition temperature ($\equiv T_{\rm BEC}(N_{\rm F}/6)$) is determined from the equation, $N_{\rm F}/6=N_{\rm B}^x=N_{\rm B}^y=N_{\rm B}^z$, which gives
\begin{equation}
T_{\rm BEC}(N_{\rm F}/6)={2\pi \over \zeta(3/2)M}
\left(
{N_{\rm F} \over 6}
\right)^{2/3}
=0.066T_{\rm F},
\label{eq.25b}
\end{equation}
where $\zeta(3/2)=2.612$ is the zeta function. 
\par
In contrast, when $\delta v^{-1}>0$, Eq. (\ref{eq.25}) shows that the $p_y$- and $p_z$-wave components no longer satisfy the BEC condition. In the extreme case when $\delta v^{-1}\gg 1$, the Bose chemical potentials $\mu_{\rm B}^{y}$ and $\mu_{\rm B}^z$ in Eq. (\ref{eq.25}) are much lower than zero, so that one can ignore the contributions of these components to the number equation (\ref{eq.24a}), as $N_{\rm F}/2=N_{\rm B}^x$. This means that most atoms form bound molecules in the $p_x$-wave Cooper channel, which is quite different from the case of $\delta v^{-1}=0$, where only one third of Fermi atoms contribute to $p_x$-wave molecules. Because of this, the BEC phase transition temperature $(\equiv T_{\rm BEC}(N_{\rm F}/2)$) in this extreme case is higher than of $\delta v^{-1}=0$ in Eq. (\ref{eq.25b}) as,
\begin{equation}
T_{\rm BEC}(N_{\rm F}/2)={2\pi \over \zeta(3/2)M}
\left(
{N_{\rm F} \over 2}
\right)^{2/3}
=0.137T_{\rm F}.
\label{eq.25c}
\end{equation}
Figure \ref{fig5}(c) shows that the BEC phase transition temperature $T_{\rm BEC} (N_{\rm B}^x)$ calculated from Eqs. (\ref{eq.24a})-(\ref{eq.25}) monotonically increases from $T_{\rm BEC}(N_{\rm F}/6)$ to $T_{\rm BEC}(N_{\rm F}/2)$, with increasing the anisotropy parameter $\delta v^{-1}$. The well agreement of $T_{\rm c}^{p_x}$ with $T_{\rm BEC}$ shown in this figure indicates that the anisotropy dependence of $T_{\rm c}^{p_x}$ in this regime comes from the increase of the molecular bosons in the $p_x$-wave Cooper channel, with increasing the uniaxial anisotropy of the $p$-wave interaction.
\par
\begin{figure}
\centerline{\includegraphics[width=8cm]{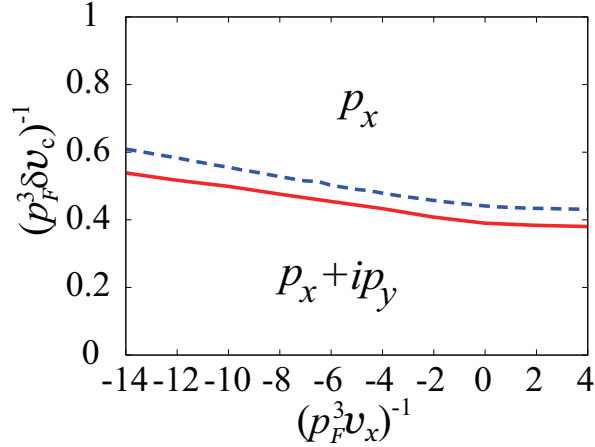}}
\caption{(Color online) Critical value $\delta v_{\rm c}^{-1}$ of the anisotropy parameter at which $T_{\rm c}^{p_x+ip_y}$ vanishes (solid line). The dashed line shows the result within the BCS-Leggett theory. In obtaining the solid line, we have taken a small but finite value of $T$ ($\lesssim 0.01T_{\rm F}$) because of computational problems.}
\label{fig6}
\end{figure}
\par
In contrast to $T_{\rm c}^{p_x}$, we see in Fig. \ref{fig5} that the $p_x+ip_y$-wave superfluid phase transition temperature $T_{\rm c}^{p_x+ip_y}$ decreases with increasing $\delta v^{-1}$, to eventually vanish at a critical value $\delta v_{\rm c}^{-1}$. (Note that this vanishing $T_{\rm c}^{p_x+ip_y}$ has already been expected in Fig. \ref{fig2}.) Evaluating this critical value $\delta v_{\rm c}^{-1}$ in the whole interaction strength, we obtain Fig. \ref{fig6}. This figure shows that $\delta v_{\rm c}^{-1}$ is not so sensitive to the interaction strength, to always lie in the narrow range, $0.4\lesssim (p_{\rm F}^3\delta v)^{-1}\lesssim 0.6$.
\par
To understand this behavior of $\delta v_{\rm c}^{-1}$, since thermal fluctuations are absent at $T=0$, it is convenient to employ the BCS-Leggett theory \cite{Leggett}, which consists of the coupled Eq. (\ref{eq.18c}) at $T_{\rm c}^{p_x+ip_y}=0$ with the mean-field number equation at $T=0$, 
\begin{equation}
N_{\rm F}=
{T \over 2}\sum_{{\bm p},i\omega_n}
{\rm Tr}
[\tau_3{\hat G}_0({\bm p},i\omega_n)]
=
{1 \over 2}
\sum_{\bm p}
\left[
1-
{\xi_{\bm p} \over \sqrt{\xi_{\bm p}^2+|\Delta_{p_x}({\bm p})|^2}}
\right].
\label{eq.25d}
\end{equation}
As shown in Fig. \ref{fig6}, the BCS-Leggett theory semi-quantitatively reproduces the TMA result for $\delta v_{\rm c}^{-1}$. In the weak-coupling BCS regime ($|\Delta_{p_x}({\bm p})|\ll\varepsilon_{\rm F}$), the number equation (\ref{eq.25d}) simply gives $\mu=\varepsilon_{\rm F}$. Substituting this into Eq. (\ref{eq.18c}) with $T_{\rm c}^{p_x+ip_y}=0$, one obtains the upper bound of $\delta v_{\rm c}^{-1}$ in the BCS-Leggett theory as
\begin{equation}
(p_{\rm F}^3\delta v_{\rm c})^{-1}={2 \over \pi} = 0.64.
\label{eq.20} 
\end{equation} 
In the strong coupling regime where the chemical potential is negative and $|\mu|\gg |\Delta_{p_x}({\bm p})|$, the BCS-Leggett theory gives the lower bound of $\delta v_{\rm c}^{-1}$ as
\begin{equation}
(p_{\rm F}^3\delta v_{\rm c})^{-1}=
{64 \over 5|k_0|}\sum_{\bm p}
{F_{n=3}^4({\bm p}) \over p^2}
+O
\left(
{\sqrt{2m|\mu|} \over |k_0|}
\right)
=
0.44+
O\left(
{\sqrt{2m|\mu|} \over |k_0|}
\right).
\label{eq.21}
\end{equation}
\par
Strictly speaking, although the TMA result for $\delta v_{\rm c}^{-1}$ coincides with Eq. (\ref{eq.20}) in the weak-coupling limit, it is still different from Eq. (\ref{eq.21}) even in the strong-coupling limit. This is because the finite value of the effective range ($k_0=-30p_{\rm F}$) causes an effective repulsive interaction between bound molecules in the latter limit, leading to the so-called quantum depletion \cite{Pethick}. Indeed, in the strong-coupling BEC regime, the last term in Eq. (\ref{eq.18e}), which describes fluctuation corrections to the mean-field Green's function ${\hat G}_0({\bm p},i\omega_n)$, modifies the mean-field number equation (\ref{eq.25d}) as
\begin{equation}
{N_{\rm F} \over 2}
=
{1 \over 4}
\sum_{\bm p}
\left[
1-
{\xi_{\bm p} \over \sqrt{\xi_{\bm p}^2+|\Delta_{p_x}({\bm p})|^2}}
\right]
+{8 \over 3\sqrt{\pi}}\sum_{i=x,y,z}
\left(
{N_{\rm F} \over 2}a_{{\rm B},i}^3
\right)^{1 \over 2},
\label{eq.22}
\end{equation}
where $a_{{\rm B},x}=(374/15)|k_0|^{-1}$, and $a_{{\rm B},y}=a_{{\rm B},z}=a_{{\rm B},x}/3$. The last term in Eq. (\ref{eq.22}) has the same form as the quantum depletion in a Bose superfluid with $N_{\rm F}/2$ bosons, when we interpret $a_{{\rm B},i}$ as an effective repulsive interaction between tightly bound $p_i$-wave molecules. When we include this quantum depletion, the lower bound in Eq. (\ref{eq.21}) is improved as 
\begin{equation}
(p_{\rm F}^3\delta v_{\rm c})^{-1}
=
{157 \over 135\pi}+
O\left(
{\sqrt{2m|\mu|} \over |k_0|}
\right)
=
0.38+
O\left(
{\sqrt{2m|\mu|} \over |k_0|}
\right),
\label{eq.23}
\end{equation}
which agrees well with the TMA result (solid line in Fig. \ref{fig6}) in the strong-coupling regime.
\par
\begin{figure}
\centerline{\includegraphics[width=8cm]{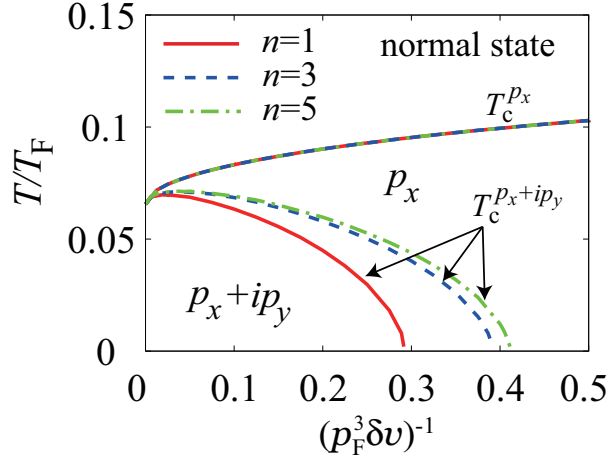}}
\caption{(Color online) Superfluid phase transition temperatures $T_{\rm c}^{p_x}$ and $T_{\rm c}^{p_x+ip_y}$ for various values of $n$ in the cutoff function $F_n({\bm p})$. We take $(p_{\rm F}^3v_x)^{-1}=0$.}
\label{fig7}
\end{figure}
\par
Before ending this section, we comment on the the cutoff function $F_{n=3}({\bm p})$ we are using.  As shown in Fig. \ref{fig7}, while the phase transition temperature $T_{\rm c}^{p_x}$ is almost independent of $n$, $T_{\rm c}^{p_x+ip_y}$ depends on this parameter. Generalizing Eq. (\ref{eq.21}) to the case with an arbitrary $n$, one finds that the lower bound (in the BCS-Leggett theory) explicitly depends on $n$ as
\begin{equation}
(p_{\rm F}^3\delta v_{\rm c})^{-1}=
{4 \over 15\pi}
\left(
3-{1 \over 2n}
\right)
\left(
2-{1 \over 2n}
\right)
+
O\left(
{\sqrt{2m|\mu|} \over |k_0|}
\right).
\label{eq.24}
\end{equation}
These $n$-dependences of $T_{\rm c}^{p_x+ip_y}$ and $\delta v_{\rm c}^{-1}$ are because the factor $p_x$ in $\Delta_{p_x}({\bm p})=p_xb_xF_n({\bm p})$ enhances this superfluid order parameter in the high momentum region, so that physical quantities in the $p_x$-wave superfluid phase depend on how this enhancement is suppressed by the cutoff function $F_n({\bm p})$. This implies that, in addition to the observable parameter set $(v_x^{-1}, \delta v^{-1}, k_0)$, one need one more experimental information about the high momentum regime of a real $p$-wave interaction, in order to unambiguously predict the phase boundary between the $p_x$-wave and $p_x+ip_y$-wave superfluid phases. We briefly note that Fig. \ref{fig7} shows that our choice ($n=3$) is close to the case of discrete cutoff (which corresponds to $n=\infty$.)
\par
\begin{figure}
\centerline{\includegraphics[width=15cm]{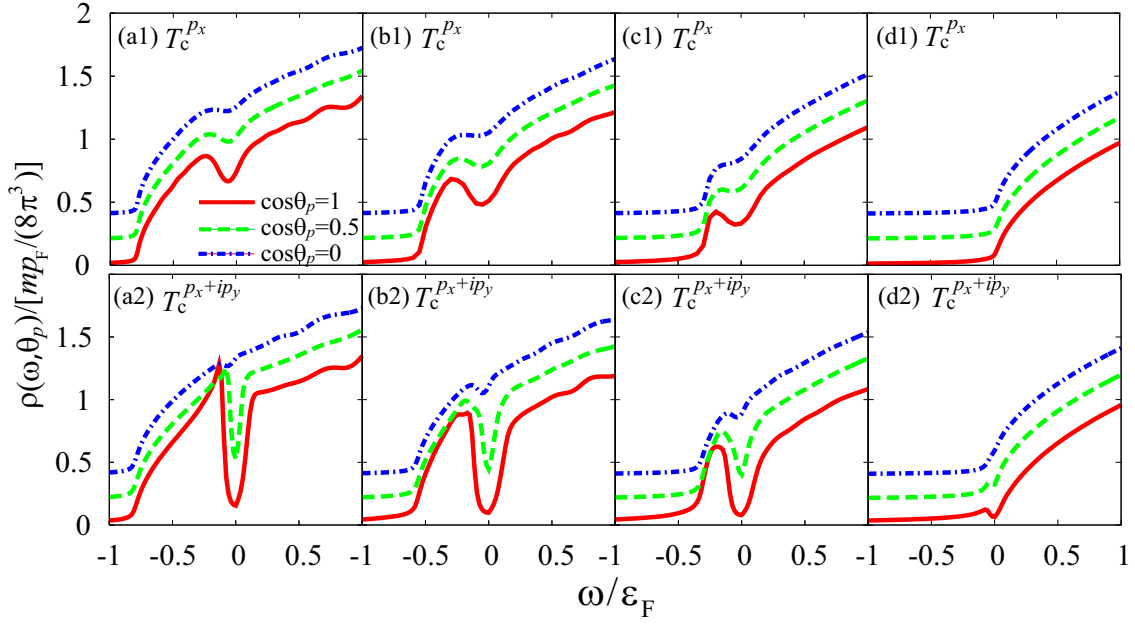}}
\caption{(Color online) Calculated angle-resolved density of states (ARDOS) $\rho(\omega,\theta_{\bm p})$. (a1)(a2) $(p_{\rm F}^3v_x)^{-1}=-12$. (b1)(b2) $(p_{\rm F}^3v_x)^{-1}=-8$. (c1)(c2) $(p_{\rm F}^3v_x)^{-1}=-4$. (d1)(d2) $(p_{\rm F}^3v_x)^{-1}=0$. Upper and lower figures show the results at $T_{\rm c}^{p_x}$ and $T_{\rm c}^{p_x+ip_y}$, respectively. In each figure, we offset the results by 0.2.
}
\label{fig8}
\end{figure}
\par
\section{Angle-resolved density of states and strong coupling effects near the phase boundaries.}
\par
Figure \ref{fig8} shows the angle resolved density of states $\rho(\omega,\theta_{\bm p})$ (ARDOS) at $T_{\rm c}^{p_x}$ (upper figures) and $T_{\rm c}^{p_x+ip_y}$ (lower figures). In Fig. \ref{fig8}(a1), a dip structure is shown around $\omega=0$. Since the superfluid order parameter vanishes at $T_{\rm c}^{p_x}$, this is just a pseudogap originating from $p$-wave pairing fluctuations. This many-body phenomenon is non-monotonic in the sense that, while this pseudogap is more remarkable in Fig. \ref{fig8}(b1), it gradually becomes obscure with further increasing the interaction strength, to eventually vanish, as shown in Figs. \ref{fig8}(c1) and (d1). Figures \ref{fig8}(a1)-(c1) also show that the pseudogap structure at $T_{\rm c}^{p_x}$ is anisotropic in momentum space, and is the most remarkable in the $p_x$ direction ($\cos\theta_{\bm p}=1$). 
\par
To simply explain this anisotropic pseudogap phenomenon, it is convenient to employ the static approximation for pairing fluctuations \cite{Levin}. Noting that the particle-particle scattering matrix $\Gamma_{x,x}^{-+}({\bm q}=0,i\nu_n=0)$ in the $p_x$-wave channel diverges at $T_{\rm c}^{p_x}$ \cite{Thouless}, we may approximate the (1,1) component of the TMA self-energy in Eq. (\ref{eq.12}) in the normal state near $T_{\rm c}^{p_x}$ to
\begin{eqnarray}
\Sigma_{11}({\bm p},i\omega_n) \simeq  
{2 \over \beta}
\sum_{{\bm q},i\nu_n}\Gamma_{x,x}^{-+}({\bm q},i\nu_n)
F_n^2({\bm p})p_x^2G^{22}_0({\bm p},i\omega_n)
\equiv
{\Delta^2_{\rm pg}({\bm p}) \over i\omega_n+\xi_{\bm p}},
\label{eq.28}
\end{eqnarray}
where  
\begin{eqnarray}
\Delta^2_{\rm pg}({\bm p})=
\left[{2 \over \beta}\sum_{{\bm q},i\nu_n}
\Gamma_{x,x}^{-+}({\bm q},i\nu_n)
\right]
p_x^2F_n^2({\bm p})
\equiv {b_{\rm pg}^x}^2p_x^2F_n^2({\bm p})
\label{eq.29}
\end{eqnarray}
is the so-called pseudogap parameter \cite{Levin,Perali}. In obtaining Eq. (\ref{eq.28}), we have only retained effects of the strongest $p_x$-wave pairing fluctuations near $T_{\rm c}^{p_x}$, and have ignored fluctuation contributions from the $p_y$-wave and $p_z$-wave Cooper channels. Substituting Eq. (\ref{eq.28}) into the (1,1) component of the TMA Green's function in Eq. (\ref{eq.10}), one obtains
\begin{eqnarray}
G_{11}({\bm p},i\omega_n)
&=&
{1 
\over
i\omega_n-\xi_{\bm p}
-
{\displaystyle \Delta^2_{\rm pg}({\bm p})
 \over \displaystyle i\omega_n+\xi_{\bm p}}
}
\nonumber
\\
&=&
-
{i\omega_n+\xi_{\bm p} 
\over 
\omega_n^2+\xi_{\bm p}^2+\Delta^2_{\rm pg}({\bm p})}.
\label{eq.30}
\end{eqnarray}
The first line in Eq. (\ref{eq.30}) means that the pseudogap parameter $\Delta_{\rm pg}({\bm p})$ works as a coupling between the particle branch $\omega=\xi_{\bm p}$ and the hole branch $\omega=-\xi_{\bm p}$.  On the viewpoint of this particle-hole coupling, the pseudogap may be interpreted as a result of the level repulsion between the particle and hole branches around $\omega=0$ \cite{Inotani,Tsuchiya1,Watanabe}.
\par
The last expression in Eq. (\ref{eq.30}) is just the same form as the diagonal component of the Green's function in the ordinary mean-field BCS theory. This coincidence immediately gives the BCS-type single-particle excitation spectra $E_{\bm p}^{\pm}=\pm \sqrt{\xi_{\bm p}^2+|\Delta_{\rm pg}({\bm p})|^2}$, having the excitation gap,
\begin{eqnarray}
\Delta E \left( \theta_p \right)
=
\left\{
\begin{array}{ll}
2|b_{\rm pg}^x\cos\theta_{\bm p}|
\sqrt{2m\mu-m^2|b_{\rm pg}^x\cos\theta_{\bm p}|^2}
& 
~~~~~(\mu \ge m|b_{\rm pg}^x\cos\theta_{\bm p}|^2), \\
2|\mu|
& ~~~~~(\mu < m|b_{\rm pg}^x\cos\theta_{\bm p}|^2),
\end{array}
\right.
\label{eq.32a}
\end{eqnarray}
where the cutoff function $F_{n=3}({\bm p})$ has been approximated to unity. (Note that $p_{\rm c}\gg p_{\rm F}$.) This gap is actually a pseudogap, when one correctly includes a finite lifetime of preformed Cooper pairs, which is ignored in the static approximation \cite{Levin}. 
\par
Equation (\ref{eq.32a}) indicates that, as expected, the anisotropic pseudogap phenomenon shown in Fig.\ref{fig8}(a1)-(c1) originates from the anisotropic $p_x$-wave pairing fluctuations, described by the pseudogap parameter $\Delta^2_{\rm pg}({\bm p})\propto p_x^2$. When $U_x=U_y=U_z$, fluctuations in all the three $p_i$-wave Cooper channels ($i=x,y,z$) are equally enhanced near the superfluid instability, so that the pseudogap parameter in Eq. (\ref{eq.29}) is replaced by the isotropic one,
\begin{eqnarray}
\Delta^2_{\rm pg}({\bm p})=
\left[{2 \over \beta}\sum_{{\bm q},i\nu_n}
\Gamma_{x,x}^{-+}({\bm q},i\nu_n)
\right]
p^2F_n^2({\bm p})
\label{eq.31}
\end{eqnarray}
where we have used the symmetry property, $\Gamma_{x,x}^{-+}=\Gamma_{y,y}^{-+}=\Gamma_{z,z}^{-+}$. The resulting pseudogap is isotropic in momentum space \cite{Inotani}.
\par
Equation (\ref{eq.32a}) also shows that the (pseudo)gap size $\Delta E(\theta_{\bm p})$ becomes isotropic in the strong coupling regime where the Fermi chemical potential is negative \cite{Ohashi,Ho}. This is because most Fermi atoms form tightly bound molecules in the strong-coupling regime, so that the threshold energy of single-particle excitations is simply dominated by the dissociation of these molecules with the binding energy $E_{\rm bind}\simeq 2|\mu|$, as in the strong-coupling BEC regime of the $s$-wave case \cite{NSR,Melo,Randeria}.
\par
The pseudogap parameter $\Delta_{\rm pg}({\bm p})$ in Eq. (\ref{eq.29}) also explains the non-monotonic behavior of the pseudogap structure in terms of the interaction strength shown in Figs. \ref{fig8}(a1)-(d1) \cite{Inotani}. Since pairing fluctuations are stronger for a stronger pairing interaction, the factor $b_{\rm pg}^x$ appearing in Eq. (\ref{eq.29}) also becomes larger, which enhances the pseudogap parameter $\Delta_{\rm pg}({\bm p})$. At the same, since strong pairing fluctuations are known to decrease the Fermi chemical potential $\mu$ \cite{Ohashi,Ho} as shown in Fig. \ref{fig3}, the effective Fermi momentum defined by ${\tilde p}_{\rm F}=\sqrt{2m\mu}$ becomes small. This decreases the pseudogap parameter at the effective Fermi momentum because $\Delta^2_{\rm pg}({\tilde {\bm p}}_{\rm F})\sim {\tilde p}^2_{{\rm F},x}\sim 2m\mu$. As a result, while the pseudogap first becomes remarkable with increasing the interaction strength in the weak-coupling region because of the enhanced pairing fluctuations, it gradually shrinks when the decrease of the Fermi chemical potential dominantly contributes to $\Delta_{\rm pg}({\bm p})$. In the case of Fig. \ref{fig8}(d1), one has $\mu(T_{\rm c}^{p_x})\simeq 0$. Thus, the low momentum region $|{\bm p}|\sim 0$ dominantly contributes to the density of states around $\omega=0$, leading to the vanishing pseudogap in this figure \cite{notePG}. 
\par
We briefly note that the vanishing pseudogap in the intermediate coupling regime ($(p_{\rm F}^3v_x)^{-1}\sim 0$) is quite different from the $s$-wave case, where the pseudogap monotonically develops, as one passes through the BCS-BEC crossover region. This is simply because the contact-type $s$-wave pairing interaction is independent on the momentum ${\bm p}$, so that the factor $p_x$ is absent in the $s$-wave pseudogap parameter.
\par
At $T_{\rm c}^{p_x+ip_y}$, since the $p_x$-wave superfluid order parameter $\Delta_{p_x}({\bm p})=b_xp_xF_{n=3}({\bm p})\propto p_x$ is already present, ARDOS $\rho(\omega,\cos\theta_{\bm p}=1, 0.5)$ in the low energy region is dominated by the superfluid energy gap, as shown in Figs. \ref{fig8}(a2)-(c2). On the other hand, such a gap structure is not shown in $\rho(\omega,\cos\theta_{\bm p}=0)$ in the weak-coupling regime (Fig. \ref{fig8}(a2)), because of the vanishing $p_x$-wave superfluid order parameter there. However, ARDOS in the nodal direction ($\cos\theta_{\bm p}=0$) gradually exhibits a dip structure around $\omega=0$, with increasing the pairing interaction. (See Figs. \ref{fig8}(b2) and (c2).) Since the $p_x+ip_y$-wave superfluid order parameter still vanishes at $T_{\rm c}^{p_x+ip_y}$, this is a pseudogap induced by fluctuations in the $p_y$- and $p_z$-wave Cooper channels. Indeed, when we apply the static approximation to the region near $T_{\rm c}^{p_x+ip_y}$, the (1,1) component of the single-particle Green's function with ${\bm p}=(0,p_y,p_z)$ is reduced to Eq. (\ref{eq.30}) where the pseudogap parameter is replaced by
\begin{equation}
\Delta^2_{\rm pg}(0,p_y,p_z)=
\sum_{i=y,z}
\left[{2 \over \beta}\sum_{{\bm q},i\nu_n}
\Gamma_{i,i}^{-+}({\bm q},i\nu_n)
\right]
p_i^2F_n^2({\bm p})
\equiv
\sum_{i=y,z}{b_{\rm pg}^i}^2p_i^2F_{n=3}^2({\bm p}).
\label{eq.32}
\end{equation}
Although Eq. (\ref{eq.32}) is similar to Eq. (\ref{eq.29}), the former involves effects of $p_x$-wave superfluid order parameter $\Delta_{p_x}({\bm p})$. Since gapless Fermi excitations only remain along the line node of the $p_x$-wave superfluid order parameter, pairing fluctuations described by $b_{\rm pg}^{i=y,z}(T=T_{\rm c}^{p_x+ip_y})$ in Eq. (\ref{eq.32}) are weaker than pairing fluctuations described by $b_{\rm pg}^x(T=T_{\rm c}^{p_x})$ in Eq. (\ref{eq.29}). This explains why the pseudogap appearing  in ARDOS $\rho(\omega,\cos\theta_{\bm p}=0)$ at $T_{\rm c}^{p_x+ip_y}$ (Figs. \ref{fig8}(b2) and (c2)) is less remarkable, compared to the dip structure in $\rho(\omega,\cos\theta_{\bm p}=1)$ at $T_{\rm c}^{p_x}$ (Figs. \ref{fig8}(a1)-(c1)).
\par
The reason for the vanishing superfluid gap and pseudogap gap in Fig. \ref{fig8}(d2) is the same as that in the case of Fig. \ref{fig8}(d1). That is, at this interaction strength, the chemical potential is very small ($\mu(T_{\rm c}^{p_x+ip_y})=0.03\varepsilon_{\rm F}$), so that the low-momentum region ($|{\bm p}|\sim 0$) dominantly contributes to ARDOS $\rho(\omega,\theta_{\bm p})$ around $\omega=0$. Thus, the $p_x$-wave superfluid order parameter $\Delta_{p_x}({\bm p})\propto p_x$, as well as effects of the pseudogap parameter $\Delta_{\rm pg}^2(0,p_y,p_z)\propto p_y^2+p_z^2$ in Eq. (\ref{eq.32}), do not almost affect ARDOS around $\omega=0$ in this case.
\par
\begin{figure}
\centerline{\includegraphics[width=15cm]{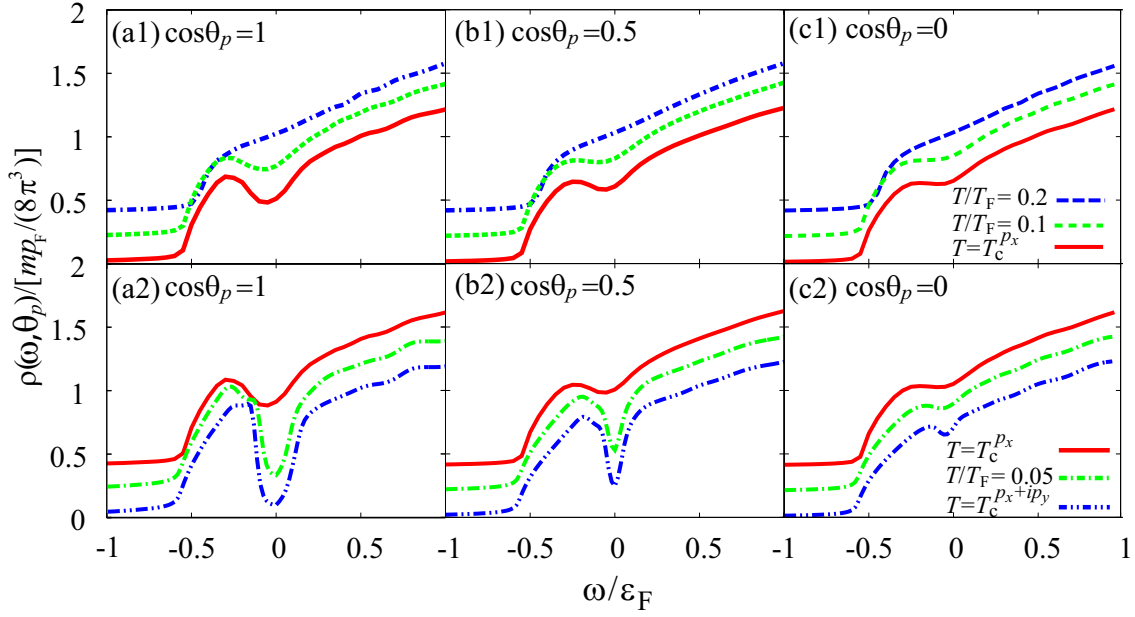}}
\caption{(Color online) Calculated angle-resolved density of states (ARDOS) $\rho(\omega,\theta_{\bm p})$ at various temperatures, when $((p_{\rm F}^3v_x)^{-1}, (p_{\rm F}^3\delta v)^{-1})=(-8, 0.3)$. (a1)(a2) $\cos\theta_{\bm p}=1$. (b1)(b2) $\cos\theta_{\bm p}=0.5$. (c1)(c2) $\cos\theta_{\bm p}=0$. The upper and lower figures show the results in the normal state, and in the $p_x$-wave superfluid state, respectively. Evaluating the pseudogap temperature $T^*(\theta_{\bm p})$ as the temperature below which a dip structure appears in $\rho(\omega,\theta_{\bm p})$ around $\omega=0$, one obtains $T^*(\cos\theta_{\bm p}=1)=0.13T_{\rm F}$, $T^*(\cos\theta_{\bm p}=0.5)=0.11T_{\rm F}$, and $T^*(\cos\theta_{\bm p}=0)=0.1T_{\rm F}$. In each figure, we offset the results by 0.3.}
\label{fig9}
\end{figure}
\par
\begin{figure}
\centerline{\includegraphics[width=10cm]{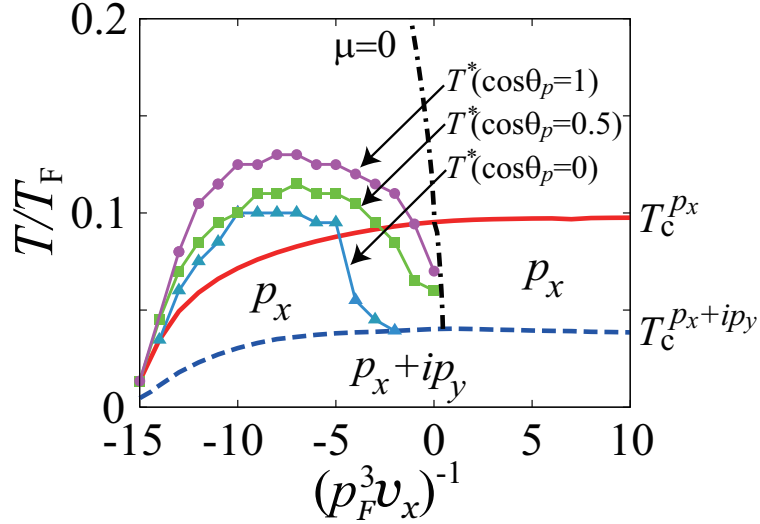}}
\caption{(Color online) Characteristic temperature $T^*(\theta_{\bm p})$ below which a dip structure appears in ARDOS $\rho(\omega,\theta_{\bm p})$. We take $(p_{\rm F}^3\delta v)^{-1}=0.3$. The dashed-dotted line shows the temperature at which the Fermi chemical potential $\mu$ vanishes. The chemical potential $\mu$ is negative in the right side of this line, so that this strong-coupling regime may be regarded as a gas of two-body bound molecules with the binding energy $E_{\rm bind}\sim 2|\mu|$ \cite{Inotani}, rather than a gas of Fermi atoms.}
\label{fig10}
\end{figure}
\par
Figures \ref{fig9}(a1) and (a2) show that the pseudogap in $\rho(\omega,\cos\theta_{\bm p}=1)$ in the normal state continuously changes to the superfluid gap, as one passes through the superfluid instability at $T_{\rm c}^{p_x}$. The same phenomenon is also shown when $\cos\theta_{\bm p}=0.5$, as shown in Figs. \ref{fig9}(b1) and (b2). On the other hand, since the $p_x$-wave superfluid order parameter vanishes when $\cos\theta_{\bm p}=0$, Figs. \ref{fig9} (c1) and (c2) show how the pseudogap in the nodal direction continues developing in the $p_x$-wave superfluid phase, to be the most remarkable at $T_{\rm c}^{p_x+ip_y}$. 
\par 
When we introduce the characteristic temperature $T^*(\theta_{\bm p})$ as the temperature below which a dip structure appears in $\rho(\omega,\cos\theta_{\bm p})$, we obtain Fig. \ref{fig10}. Because $\Delta_{p_x}({\bm p})=0$ in the nodal direction ($\cos\theta_{\bm p}=0$), we may regard $T^*(\cos\theta_{\bm p}=0)$ as the pseudogap temperature \cite{Inotani,Tsuchiya1,Watanabe} in this momentum direction, below which strong pairing fluctuations induce a pseudogap in ARDOS $\rho(\omega,\cos\theta_{\bm p}=0)$. In this case, one may call the region surrounded by $T^*(\cos\theta_{\bm p}=0)$ and $T_{\rm c}^{p_x+ip_y}$ the pseudogap regime.
\par
\begin{figure}
\centerline{\includegraphics[width=8cm]{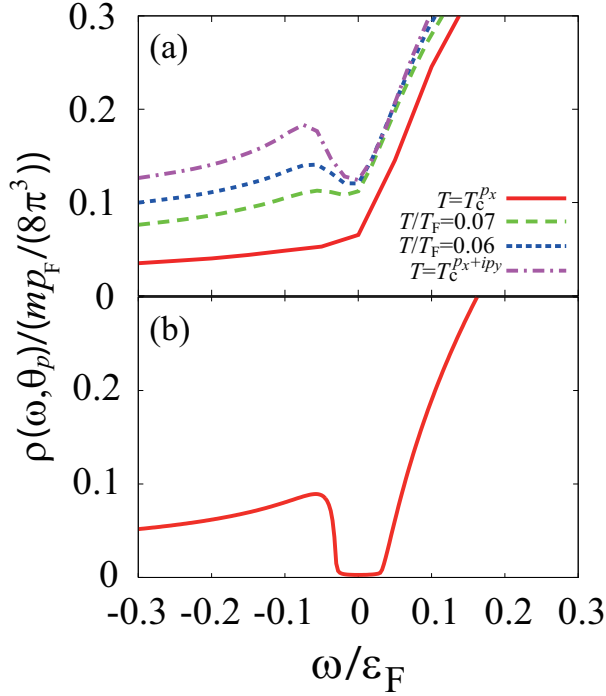}}
\caption{(Color online) (a) Angle-resolved density of states $\rho(\omega,\cos\theta_{\bm p}=1)$ in the $p_x$-wave superfluid phase. We take $((p_{\rm F}^3v_x)^{-1}, (p_{\rm F}^3\delta v)^{-1})=(0, 0.3)$. Each line is offset by 0.03. (b) Mean-field result at $T=T_{\rm c}^{p_x+ip_y}$, which is obtained by ignoring the self-energy correction ${\hat \Sigma}({\bm p},i\omega_n)$ in Eq. (\ref{eq.10}) in calculating ARDOS.
}
\label{fig11}
\end{figure}
\par
In the case of $\cos\theta_{\bm p}\ne 0$, $T^*(\theta_{\bm p})$ also has the meaning of the pseudogap temperature, when $T^*(\theta_{\bm p})>T_{\rm c}^{p_x}$. On the other hand, $T^*(\cos\theta_{\bm p}=1,~0.5)$ is lower than $T_{\rm c}^{p_x}$ around $(p_{\rm F}^3v_x)^{-1}=0$ in Fig.\ref{fig10}, which means that the superfluid gap does not appear in ARDOS when $T^*(\theta_{\bm p})\le T\le T_{\rm c}^{p_x}$. As mentioned previously, since $|\mu|\ll \varepsilon_{\rm F}$ in this intermediate coupling regime, single-particle excitations around ${\bm p}=0$ dominantly contribute to ARDOS around $\omega=0$. Because of this, a small superfluid excitation gap by a small $p_x$-wave superfluid order parameter, $\Delta_{p_x}({\bm p})\sim b_x\sqrt{2m\mu}\sim 0$, around ${\bm p}=0$ is easily smeared out by strong pairing fluctuations existing in this regime even below $T_{\rm c}^{p_x}$. Since this strong-coupling effect is gradually suppressed below $T_{\rm c}^{p_x}$, ARDOS starts to exhibit a superfluid gap structure below $T^*(\theta_{\bm p})$ (See Fig. \ref{fig11}(a).), to approach the BCS-type superfluid density of states shown in Fig. \ref{fig11}(b). Thus, $T^*(\theta_{\bm p})$ in this regime may be regarded as the characteristic temperature, below which the $p_x$-wave superfluid order overwhelms pairing fluctuations. 
\par
In Fig. \ref{fig10}, we also plot the temperature at which the Fermi chemical potential $\mu$ changes its sign \cite{noteM}. As mentioned previously, in the strong-coupling regime where $\mu<0$, since the system is dominated by tightly bound molecules, the $p$-wave character of Cooper pairs is less important. In the normal state near $T_{\rm c}^{p_x}$, this fact gives the isotropic pseudogap size $\Delta E(\theta_{\bm p})$ in Eq. (\ref{eq.32a}). In the $p_x$-wave superfluid phase, the Bogoliubov single particle excitation spectrum in the strong coupling regime,
\begin{equation}
E_{\bm p}=\sqrt{(\varepsilon_{\bm p}+|\mu|)^2+|\Delta_{p_x}({\bm p})|^2},
\label{eq.F}
\end{equation}
also has the isotropic energy gap $2|\mu|$, reflecting that the threshold energy of Fermi excitations is simply dominated by the binding energy ($E_{\rm bind}\sim 2|\mu|$) of a two-body bound molecule. Thus, the $p$-wave anisotropy is not important in the right side of this line in Fig. \ref{fig10}, as far as we consider low-energy single-particle excitations.
\par
\section{Summary}
\par
To summarize, we have discussed strong-coupling properties of a one-component superfluid Fermi gas with a uniaxially anisotropic $p$-wave pairing interaction ($U_x>U_y=U_z$). Including $p$-wave pairing fluctuations within a $T$-matrix approximation, we determined the two superfluid phase transition temperatures $T_{\rm c}^{p_x}$, which gives the phase boundary between the normal state and the $p_x$-wave superfluid state, and $T_{\rm c}^{p_x+ip_y}$ ($<T_{\rm c}^{p_x}$), which gives the phase boundary between the $p_x$-wave and $p_x+ip_y$-wave superfluid states. 
\par
We examined single-particle excitations near $T_{\rm c}^{p_x}$, as well as near $T_{\rm c}^{p_x+ip_y}$. In the normal state near $T_{\rm c}^{p_x}$, we showed that strong pairing fluctuations in the $p_x$-wave Cooper channel induce an anisotropic pseudogap phenomenon where a pseudogap structure in the angle-resolved density of states (ARDOS) is the most remarkable in the $p_x$ direction. We also showed that this pseudogap continuously changes to the $p_x$-wave superfluid gap below $T_{\rm c}^{p_x}$. On the other hand, the pseudogap was found to continue developing below $T_{\rm c}^{p_x}$ in the nodal direction ($\perp p_x$) of the $p_x$-wave superfluid order parameter, to be the most remarkable at $T_{\rm c}^{p_x+ip_y}$. Since pairing fluctuations are simply suppressed in an isotropic $s$-wave superfluid state, this phenomenon is characteristic of a $p$-wave Fermi superfluid with a nodal superfluid order parameter and with plural superfluid phases. To characterize the anisotropic pseudogap phenomenon, we determined the characteristic temperature $T^*(\theta_{\bm p})$, below which a dip structure appears in ARDOS.
\par
In this paper, we have considered the normal state, as well as the $p_x$-wave superfluid phase. To obtain the complete understanding of a $p$-wave superfluid Fermi gas with a uniaxially anisotropic $p$-wave interaction, we need to also examine the $p_x+ip_y$-wave superfluid phase below $T_{\rm c}^{p_x+ip_y}$. In addition, for simplicity, we employed the BCS Hamiltonian with a $p$-wave pairing interaction, which implicitly assumes a broad Feshbach resonance. In this regard, all current experiments are using a narrow $p$-wave Feshbach resonance, so that it is an important problem to clarify how the resonance width affects strong-coupling properties of a $p$-wave superfluid Fermi gas. Since a $p$-wave superfluid state is known to be sensitive to spatial inhomogeneity, inclusion of a realistic harmonic trap also remains as our future problem. 
\par
The realization of a $p$-wave superfluid Fermi gas is an exciting challenge, in order to qualitatively go beyond the current stage of cold Fermi gas physics that the $s$-wave Fermi superfluid has only been realized. Since the anisotropic pairing is a crucial key in a $p$-wave Fermi superfluid, our results would contribute to understanding how this character affects many-body properties of an ultracold Fermi gas, especially in the superfluid phase. 
\par
\acknowledgments
We would like to thank M. Sigrist, S. Tsuchiya, S. Watabe, R. Watanabe and T. Kashimura for useful discussions. This work was supported by KiPAS project in Keio University. YO was also supported by Grant-in-Aid for Scientific research from MEXT and JSPS in Japan (25105511, 25400418, 15H00840).
\par


\begin{thebibliography}{99}
\bibitem{note2} Since we are taking a finite value of the momentum cutoff $p_{\rm c}=27p_{\rm F}$, the ``strong-coupling limit" in this paper means the region where $(p_{\rm F}^3v_x)^{-1}\gg 1$, but (1) the molecular binding energy, (2) the absolute value of the chemical potential $|\mu|$, as well as (3) the superfluid order parameter, are still much smaller than the cutoff-energy $p_{\rm c}^2/(2m)$, to avoid any unphysical effect coming from this high-energy cutoff.
\bibitem{notePG} We note that, in the strong-coupling region ($\mu < 0$), the pseudogap opens again with the size $2|\mu| \simeq E_{\rm bind}$, as in the strong-coupling region of $s$-wave case, although we do not explicitly show the result here.
\bibitem{noteM} Of course, this line also exists in the $p_x+ip_y$-wave superfluid phase below $T_{\rm c}^{p_x+ip_y}$. However, since we have not calculated $\mu$ in this region, Fig, \ref{fig10} only shows $\mu(T\ge T_{\rm c}^{p_x+ip_y})$.
\end{thebibliography}
\end{document}